\documentclass[showpacs, pra,onecolumn,preprintnumbers ,amsmath, amssymb, superscriptaddress, aps]{revtex4-2}
\usepackage{amsfonts}
\usepackage{color}
\usepackage{amsmath,amssymb}
\usepackage{pifont}
\usepackage{amssymb}  
\usepackage{bbold}
\usepackage{float}
\usepackage{subfloat}

\usepackage[caption=false]{subfig}
\usepackage{tikz}
\usepackage{makecell}
\usepackage{subfig}
\usepackage{pifont}   
\usepackage{graphicx} 
\graphicspath{{Figures/}}
\usepackage{dcolumn}  
\usepackage{bm}       
\usepackage{multirow} 
\usepackage{placeins}
\usepackage[colorlinks]{hyperref}
\usepackage{mathtools}
\usepackage{appendix}

\def \be{\begin{align}}
	\def \ee{\end{align}}
\def \bea{\begin{eqnarray}}
	\def \eea{\end{eqnarray}}

\begin{document}
	\title{Boosting energy levels in graphene magnetic quantum dots through  magnetic flux and inhomogeneous gap}
	
	
	\author{Mohammed El Azar}
		\email{elazar.m@ucd.ac.ma}
	\affiliation{ Laboratory of Theoretical Physics, Faculty of Sciences, Choua\"ib Doukkali University, PO Box 20, 24000 El Jadida, Morocco}
	\author{Ahmed Bouhlal}
	\affiliation{ Laboratory of Theoretical Physics, Faculty of Sciences, Choua\"ib Doukkali University, PO Box 20, 24000 El Jadida, Morocco}
	\author{Ahmed Jellal}
	\email{a.jellal@ucd.ac.ma}
	\affiliation{ Laboratory of Theoretical Physics, Faculty of Sciences, Choua\"ib Doukkali University, PO Box 20, 24000 El Jadida, Morocco}
	\affiliation{
		Canadian Quantum  Research Center,
		204-3002 32 Ave Vernon, BC V1T 2L7,  Canada}

	\begin{abstract}
		
		We study the effects of a magnetic flux and an inhomogeneous gap on the energy spectrum of graphene magnetic quantum dots (GMQDs). By considering the Dirac equation in the infinite mass framework, we can analytically obtain eigenspinor expressions. By applying boundary conditions, we obtain an energy spectrum equation in terms of system parameters such as radius, magnetic field, energy, flux, and gap. In the infinite limit, we recover Landau levels for graphene in a magnetic field. We show that the energy spectrum increases significantly in the presence of flux and a gap inside the GMQDs, which prolongs the lifetime of the trapped electron states. We show that higher flux also produces new Landau levels of negative angular momentum. Meanwhile, we find that the gap increases the separation between the electron and hole energy bands. As shown in the radial probability analysis, flux and gap emerge as influential factors in controlling electron mobility, affecting confinement, and prolonging the presence of quasi-bound states.


	\vspace{0.25cm}
	
\end{abstract}
\maketitle

\section{Introduction}
Graphene is a 2D material made entirely of carbon atoms. {It has no gap between the valence and conduction bands and exhibits a linear band structure near the Fermi level  \cite{sprinkle2009first,zhan2012engineering}.
This behavior is similar to that of relativistic massless particles described by the Dirac-Weyl equation \cite{katsnelson2007graphene,neto2016two}.
} 
Graphene possesses remarkable physical properties such as record thermal conductivity and very high electron mobility, making it one of the strongest materials \cite{katsnelson2007graphene, Xia10}. 
Recently, graphene has been a much sought-after substance in scientific studies due to its unique electrical characteristics \cite{neto2009electronic,peres2010colloquium,Repetsky18}, including  Klein tunneling \cite{katsnelson2006chiral,gutierrez2016klein}, quantum Hall effect \cite{novoselov2005two,jiang2007quantum,barlas2012quantum,myoung2019splitting} and Hofstadter butterfly spectrum \cite{zhang2008tuning,kooi2018genesis}, and more.
{On the other hand,} quantum dots (QDs) made of graphene are tiny particles that combine the properties of both carbon dots and graphene \cite{sun2013recent}. They have wave functions that are confined within a small disk of radius $R$ due to the effects of edge and quantum confinement. As a result, QDs have unique optical, electrical, spin, and photoelectric properties, making them a highly versatile and multifunctional material. There are two methods for making graphene QDs: a top-down method, in which sheets of graphene are cut, and a bottom-up method, in which large graphene-like molecules with a well-defined chemical structure are synthesized \cite{simpson2002synthesis,wu2004branched}. 

Many studies have investigated transport in graphene using electrostatic QDs \cite{bardarson2009electrostatic,lee2016imaging,freitag2018large}. However, electrically controlled QDs can make it difficult to apply local gates, which are common in two-dimensional electron gas systems, due to the Klein tunneling effect. Nevertheless, electron mobility at the contact can be confined in the presence of a magnetic field \cite{ghosh2008conductance,oroszlany2008theory,taychatanapat2015conductance,pena2022electron}. Researchers have shown that localized states with discrete energy spectra can exist inside a magnetically controlled QD. A magnetic field can locally create such a QD by shielding a uniform magnetic field \cite{sim1998magnetic}. The use of magnetically controlled QDs makes it possible to circumvent the problem of Klein tunneling \cite{de2007magnetic,myoung2009tunneling,esmailpour2018effect}.
Numerous experimental and theoretical studies have investigated the energy spectrum of QDs when subjected to  magnetic fields \cite{schnez2008analytic, schnez2009observation, recher2009bound, giavaras2009magnetic, grujic2011electronic, zarenia2011energy,Farsi21, Belokda23,orozco2019enhancing}. 
The confinement of electrons within QDs, especially in the context of graphene, has been the focus of extensive research \cite{Grushevskaya21, Pena2022, el2024electrons}. This research is critical to overcoming various challenges associated with controlling the electron behavior in this valuable material. However, Klein tunneling prevents the trapping of electrons falling vertically onto graphene QDs. Despite this obstacle, several studies have discovered brief trapping of electrons in QDs, prompting ongoing efforts by researchers to increase the duration of such trapping events.

We study the effects of a magnetic flux and an inhomogeneous gap on the energy spectrum of a graphene magnetic quantum dot (GMQD), where confinement is created by an applied magnetic field. We use an infinite mass boundary condition at the edge of the GMQD because it has some advantages. Indeed, it provides a simple, analytically solvable model of the confinement, even though physically the mass at the edge is not actually infinite. Mathematically, it requires the wavefunction to vanish at the boundary of the dot, mimicking total reflection. This is easier to work with than other boundary conditions, which give Dirac-like eigenstates confined inside the dot. Outside the boundary, the wavefunction unphysically explodes, but is not considered. We apply the boundary condition to obtain an analytical expression describing the energy spectrum in terms of the GMQD radius, magnetic field, magnetic flux, and energy gap. In the infinite limit, we recover the Landau levels of Dirac fermions in graphene subjected to a magnetic field.
Our analysis shows that the energy spectrum increases significantly with the magnetic flux and the energy gap within the GMQD, prolonging the lifetime of the trapped electron states. We find the emergence of new Landau levels corresponding to negative angular momentum with increasing magnetic flux. Conversely, the introduction of an energy gap widens the band gap between the electron and hole energy bands. In addition, we study the radial probability associated with our system and find oscillatory behavior. We show that the flux and the gap affect the behavior of the radial probability. As a result, we show that the magnetic flux can regulate the behavior of the electrons in the GMQD by influencing their confinement and extending the duration of quasi-bound states.


{Experimentally, we believe that the optical spectroscopy techniques provide valuable insight into the electronic spectrum of graphene magnetic quantum dots (GMQDs) \cite{Bockelmann96, Dekel98, Lorke2000}. To perform such measurements, it is necessary to fabricate a sample of graphene quantum dots and establish an appropriate magnetic field configuration. The sample should be designed to allow optical access, which can be achieved by depositing the quantum dots on a transparent substrate or embedding them in a matrix with optical properties conducive to spectroscopic studies. To perform optical spectroscopy, an appropriate setup, such as photoluminescence (PL) or Raman spectroscopy, is required. This involves configuring the system with appropriate excitation sources, detectors, and spectral analysis capabilities. The setup should allow precise control of the incident light and efficient collection of the emitted or scattered light from the sample. The GMQDs are excited by irradiation with photons of appropriate energy provided by the excitation source. In PL spectroscopy, laser light is used to illuminate the sample, while in Raman spectroscopy, monochromatic light is used to probe the vibrational modes. The light emitted or scattered by the sample is measured by detectors built into the spectroscopy setup. Spectra are recorded as a function of various parameters, such as excitation energy, magnetic field strength, or other relevant factors, to obtain comprehensive information about the electronic spectrum of the GMQDs. By following these procedures, the capabilities of optical spectroscopy can be used to gain valuable insight into the electronic properties of the GMQDs.}

The paper is organized as follows. In Sec. \ref{theory}, we present a theoretical model that provides a detailed description of our system. In Sec. \ref{calculation}, we apply a mathematical method based on hepergeometric functions to determine the eigenspinor solutions of the Dirac equation. Then, under the infinite mass boundary condition, we derive an equation governing the corresponding energy spectrum. In Sec. \ref{res}, we present our numerical results for various physical parameters of the system, including the magnetic flux, the GMQD radius, the energy gap, and the magnetic field strength. Finally, we provide a brief conclusion in Sec. \ref{cc}.

\section{Theoretical model}\label{theory}

 \begin{figure}[H]
 	\centering 
 	\includegraphics[scale=0.4]{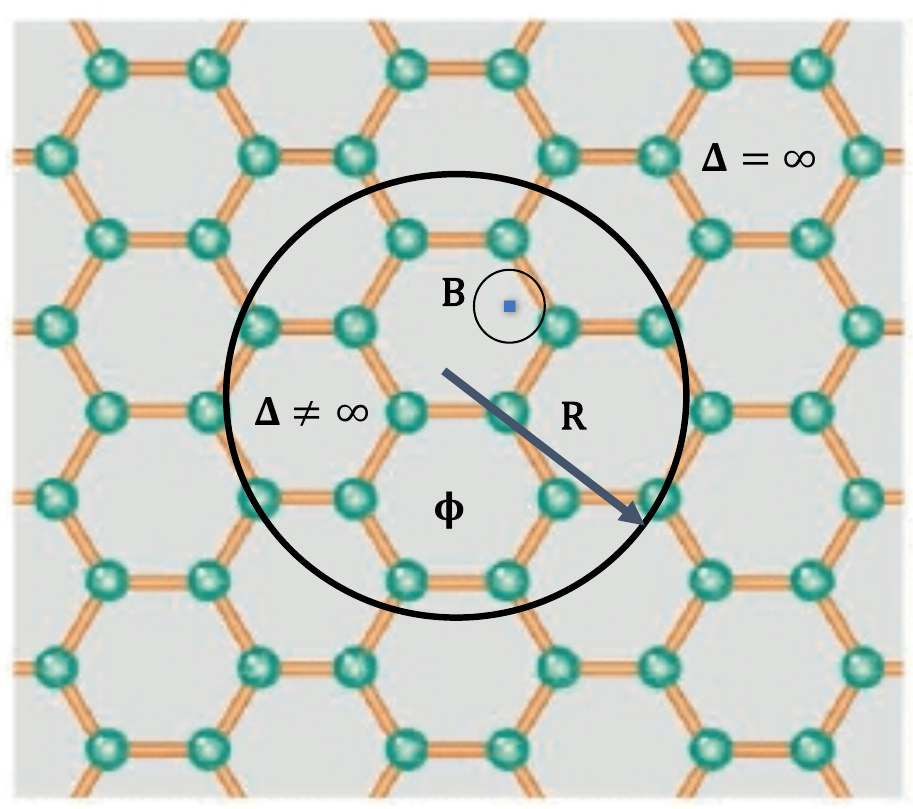}	
 	\caption{(color online) A graphene quantum dot with a radius $R$ is positioned in the horizontal plane $xy$ and exposed to a magnetic field perpendicular to the plane. Additionally, there is a magnetic flux, and  an inhomogeneous gap, with  $\Delta \ne \infty$ inside and $\Delta = \infty$ outside the dot.
 	}\label{figsystem}
 \end{figure} 
We consider a graphene magnetic quantum dot (GMQD) subjected to  an inhomogeneous gap and magnetic flux, as depicted in Fig. \ref{figsystem}. The GMQD is modeled as a circular region of graphene with radius $R$, where massless Dirac fermions are confined. An orthogonal magnetic field $B$ is applied, producing Landau quantization of energy levels. The magnetic flux threading the GMQD area is represented by $\phi$. This induces an Aharonov-Bohm effect, modifying the energy spectrum. An inhomogeneous gap $\Delta$ is introduced, such as by proximity to a substrate. This opens an energy gap in the graphene. The Dirac-Weyl equation is solved within the GMQD boundary using appropriate boundary conditions. This yields analytical expressions for the energy eigenvalues. Parameters like $R, B, \phi$, and $\Delta$ can then be varied to investigate their effects on the confined electron states and energy spectrum. Tools like probability density distributions may be analyzed to study how magnetic factors influence electron confinement and quasi-bound state lifetimes. The goal is to understand how manipulating external fields can help control electronic properties like carrier mobility in these nanostructured graphene systems.

To conduct a theoretical analysis of the present system, we construct a Hamiltonian that incorporates all the mentioned constraints. It can be expressed as follows:
\begin{equation}\label{Hamilt}
H^\tau =v_F  \vec \sigma\cdot (\vec p +e \vec A)  + \tau\Delta\sigma_z
\end{equation}
where $v_F = 10^6$ m/s represents the Fermi velocity, and $(\sigma_x, \sigma_y, \sigma_z)$ denote Pauli matrices. The momentum operator is $\vec{p}=(p_x,p_y)$, $(-e)$ represents the electron charge, and the index $\tau$ takes values +1 and -1 to distinguish between the two valleys $K$ and $K'$. The energy gap $\Delta$ and the vector potential $\vec{A_\theta}=\vec{A_1}+\vec{A_2}$ \cite{ikhdair2015nonrelativistic} are defined as
\begin{equation}
\Delta= \begin{cases}\hbar v_F \delta, & r<R \\ \infty, & \text { otherwise }\end{cases},\qquad \vec{A_1}= \begin{cases}\frac{Br}{2}\vec{e}_\theta, & r<R \\ 0, & \text { otherwise }\end{cases},\qquad \vec{A_2}= \begin{cases}\frac{\phi}{2\pi r}\vec{e}_\theta, & r<R \\ 0, & \text { otherwise }\end{cases}
\end{equation}
with $ \vec{e}_\theta$ is an unit vector in polar representation.
We may express the Hamiltonian in polar coordinates because of the spherical symmetry, knowing that
\begin{equation}
\sigma_r = \begin{pmatrix} 0 &\ e^{-i\theta}\\\\ \ e^{i\theta} &0 \\
\end{pmatrix}, \quad 
\sigma_\theta = \
\begin{pmatrix} 0 &\ -ie^{-i\theta}\\\\ \ ie^{i\theta} &0 \\
\end{pmatrix},\quad 
\sigma_z = \begin{pmatrix} 1 &\ 0\\\\ \ 0 &-1 \\
\end{pmatrix}
\end{equation}
and consequently, we can write \eqref{Hamilt} as
\begin{equation}
H^\tau =\hbar v_F \begin{pmatrix} 
	\tau\delta & \pi_+ \\
	  \pi_-  &-\tau\delta 
	  \end{pmatrix}
\end{equation}
where the operators  $\pi_\pm$ are given by 
\begin{equation}
\pi \pm=  -i e^{\mp i\theta}\left(\partial_r \mp \frac{i}{r}\partial_\theta \pm \frac{e Br}{2}\pm \frac{e\phi}{2\pi r}\right).
\end{equation}

Given that the total angular momentum operator $J_z= -i\hbar \partial_ \theta +\frac{\hbar}{2}\sigma_z $ commutes with the Hamiltonian \eqref{Hamilt} (i.e., $ [H^\tau, J_z] = 0 $), the eigenstates $ \Psi^\tau(r,\theta)=\dbinom{\Psi_A^\tau(r,\theta) }{\Psi_B^\tau(r,\theta)}$ can be categorized into the following pairs:
\begin{equation}\label{ansatz}
\Psi^\tau(r,\theta)= 
 \dbinom{e^{im\theta} \Omega_A^\tau(r) }{i e^{i(m+1)\theta} \Omega_B^\tau(r)}  
\end{equation}
where the quantum number $m \in \mathbb{Z}$ represents the eigenvalues of $J_z$. The eigenvalue equation $H^\tau \Psi^\tau=E \Psi^\tau$ leads to two coupled equations
\begin{subequations}\label{6}
	\begin{align}
	&	-\partial_r \Omega_A^\tau(r)+ \left( \frac{m+\eta}{r} +\frac{r}{2l_B^2}\right) \Omega_A^\tau(r) = (\epsilon-\tau\delta)\Omega_B^\tau(r) \label{6a}\\
	&	\partial_r \Omega_B^\tau(r)+ \left( \frac{m+1+\eta}{r}+ \frac{r}{2 l_B^2} \right) \Omega_B^\tau(r) =(\epsilon+\tau\delta)\Omega_A^\tau(r) \label{6b}		
	\end{align}
\end{subequations}	
where we have introduced the dimensionless quantities $\eta=\phi/\phi_0$ and $\epsilon=E/\hbar v_F$, with the flux unit $\phi_0=h/e$ and the magnetic length
$l_B=\sqrt{\hbar/e B}$. Substituting  \eqref{6b} into \eqref{6a}, we get   a second-order differential equation for $\Omega_A^\tau(r)$
\begin{equation}\label{e:7}
\partial_r^2\Omega_A^\tau(r)+\frac{1}{r}\partial_r \Omega_A^\tau(r)-\left( \frac{(m+\eta)^2}{r^2} +\frac{m+\eta+1}{l_B^2}+\frac{r^2}{4 l_B^4}-q^2\right) \Omega_A^\tau(r) =0
\end{equation}
and $ q=\sqrt{\epsilon^2 -\delta^2}$ is the wave number. Next, we will employ a technical method to solve the last equation. Using this approach, we will establish an equation that governs the energy levels associated with the present system.

\section{Energy spectrum}\label{calculation}

To find the solution for \eqref{e:7}, we will follow a method employed in the literature \cite{falaye2015formula}, which is based on solving a Schrödinger-type differential equation for the function $\Theta(\zeta)$. This is 
\begin{equation}\label{e:8}
\partial_\zeta^2\Theta(\zeta)+ \frac{p_1-p_2\zeta}{\zeta-p_3 \zeta^2} \partial_\zeta\Theta(\zeta)+ \frac{a\zeta^2 +b\zeta +c}{\zeta^2(1-p_3\zeta)^2} \Theta(\zeta)=0
\end{equation}
where $a, b, c$ and $p_i$ $(i=1,2,3)$ are constant parameters. This equation admits a solution in terms of the hypergeometric function $\prescript{}{2}{F}_1^{}$ and then we have
%
%
\begin{equation}\label{e:9}
\Theta(\zeta)=N \zeta^{p_4} (1-p_3\zeta)^{p_5} \prescript{}{2}{F}_1^{}\left(-\mu,\mu+2(p_4+p_5)+\frac{p_2}{p_3}-1;2p_4+p_1,p_3\zeta\right)
\end{equation}
where the two parameters $p_4$ and $p_5$ are given by
\begin{align}	
	&\label{e:1000}
	p_4=\frac{(1+p_1)+ \sqrt{(1+p_1)^2 -4c}}{2}\\
	&\label{e:100}
	p_5=\frac{b-p_2 p_4-\mu p_2}{2 p_4 + p_1+ 2\mu}
\end{align}
and  $N$ being a normalization constant.
Before delving into the applications of these formulas, let's first consider a special case where $p_3 \to 0$. In this scenario, the corresponding wave function $ \Theta(\zeta)$ can be determined as follows:
\begin{equation}\label{e:10}
 \Theta(\zeta)=N \zeta^{p_4} e^{-p_5\zeta} \prescript{}{1}{F}_1^{}\left(-\mu,2p_4+p_1,2(p_5+p_2)\zeta\right).
\end{equation}
To establish a connection with \eqref{e:8}, let's introduce the variable change $\zeta=r^2$. After some algebraic manipulations, we demonstrate that \eqref{e:7} can transform into:
\begin{equation}\label{e:11}
\partial_\zeta^2\Omega_A^\tau(\zeta)+\frac{1}{\zeta}\partial_\zeta \Omega_A^\tau(\zeta)-\frac{1}{4\zeta}\left( \frac{(m+\eta)^2}{\zeta} +\frac{m+\eta+1}{l_B^2}+\frac{\zeta}{4 l_B^4}-q^2\right) \Omega_A^\tau(\zeta) =0
\end{equation}	
As a result, it becomes clear that \eqref{e:11} is basically equal to \eqref{e:8} when we identify the related parameters in the following manner:
\begin{align}
	 p_1=1,\quad p_2=p_3=0,\quad a=-\frac{1}{16 l_B^4},\quad b=\frac{q^2}{4}-\frac{m+\eta+1}{4 l_B^2},\quad c=-\frac{(m+\eta)^2}{4}.
\end{align}
Therefore, utilizing \eqref{e:10}, we can arrive at the  solution
\begin{equation}\label{e:12}
\Omega_A^\tau(\zeta)=N \zeta^{\frac{\alpha-1}{2}} e^{\frac{-\zeta}{4 l_B^2}} \prescript{}{1}{F}_1^{}\left(\alpha-\frac{q^2 l_B^2}{2},\alpha,\frac{\zeta}{2 l_B^2}\right)
\end{equation}
where we have set the parameter $\alpha=m+\eta+1$. 
By utilizing the connection between confluent hypergeometric functions and Laguerre polynomial functions,  \eqref{e:12} transforms into:
\begin{equation}\label{e:13}
\Omega_A^\tau(\zeta)=C \zeta^{\frac{\alpha-1}{2}} e^{\frac{-\zeta}{4 l_B^2}} L\left(\frac{q^2 l_B^2}{2}-\alpha,\alpha-1,\frac{\zeta}{2 l_B^2}\right).
\end{equation}
For the second spinor component, we can use \eqref{6a} along with \eqref{e:13} to arrive at
\begin{equation}\label{e:14}
\Omega_B^\tau(\zeta)=\frac{C}{(\epsilon-\tau \delta) l_B^2} \zeta^{\frac{\alpha}{2}} e^{\frac{-\zeta}{4 l_B^2}} \left[ L\left(\frac{q^2 l_B^2}{2}-\alpha,\alpha-1,\frac{\zeta}{2 l_B^2}\right)+L\left(\frac{q^2 l_B^2}{2}-\alpha-1,\alpha,\frac{\zeta}{2 l_B^2}\right)\right]. 
\end{equation}

We have successfully determined the eigenspinors \eqref{ansatz} by completing the calculation for both spinor components. The subsequent step involves deriving the corresponding eigenvalues. However, obtaining them directly is challenging due to the complexity of the system. To overcome such a challenge, we proceed by using the ratio $\frac{\Psi_B^\tau(R,\theta)}{\Psi_A^\tau(R,\theta)}=i\tau e^{i\theta}$ at the boundary condition $r=R$ \cite{berry1987neutrino}. This process yields 
%
%
%
%
\begin{equation}\label{e:15}
\left(1-\frac{(\tau \epsilon-\delta) l_B }{\rho}\right) L\left(\frac{q^2 l_B^2}{2}-\alpha,\alpha-1,\frac{\rho^2}{2}\right)+L\left(\frac{q^2 l_B^2}{2}-\alpha-1,\alpha,\frac{\rho^2}{2}\right) =0
\end{equation}
where we have defined the parameter  $\rho=\frac{R}{l_B}$ as a normalized radius of the GMQD. 
Exploring the limit $\rho\to \infty$ is intriguing as it can provide additional insights into the characteristics of the present system. In this case, the Laguerre  function can be approximated as
%
\begin{equation}
L(m, n, z)=\frac{(m+n) !}{n ! m !} \frac{\Gamma(n+1)}{\Gamma(-m)} e^z z^{-(m+n+1)}\left[1+\mathcal{O}\left(\frac{1}{|z|}\right)\right].
\end{equation}
By manipulating algebraically \eqref{e:15} and utilizing the gamma function relation $j\Gamma(j)=\Gamma(j+1)$, we obtain an analytical formula 
\begin{equation}\label{e:18}
E_{m,\eta}=\pm \sqrt{2e\hbar^2 v_F^2 B(m+\eta +1)+\Delta^2}.
\end{equation}
This actually describes the Landau levels of graphene in the presence of an external magnetic field $B$, a magnetic flux $\phi$, and a  energy gap $\Delta$.

\section{RESULTS AND DISCUSSIONS}\label{res}

\begin{figure}[H]
	\centering
	\includegraphics[scale=0.46]{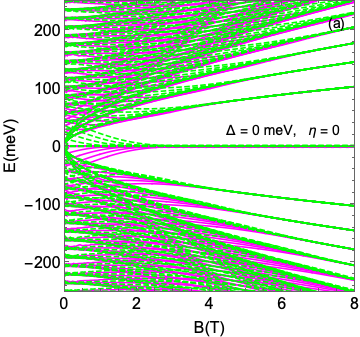}\includegraphics[scale=0.46]{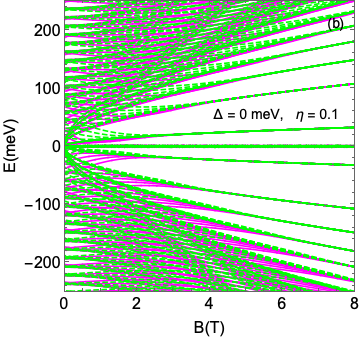}\includegraphics[scale=0.46]{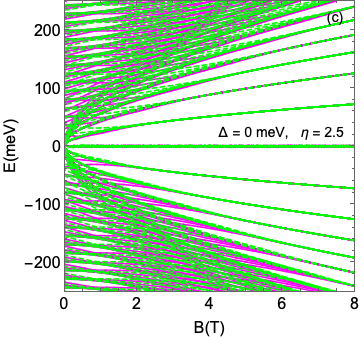}
	\includegraphics[scale=0.46]{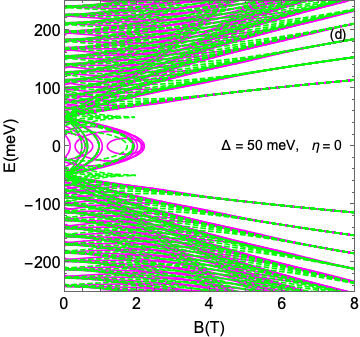}\includegraphics[scale=0.46]{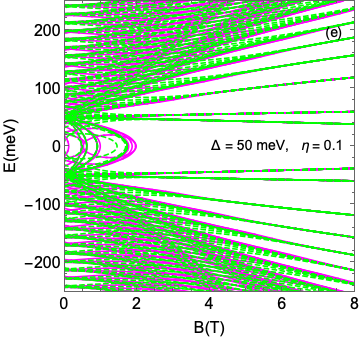}\includegraphics[scale=0.46]{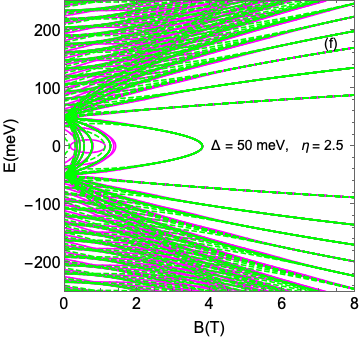}
	\caption{(color online) Energy spectrum $E$ as a function of the magnetic field $B$ for the quantum numbers $m = 0, \pm1,\cdots,\pm4$, radius $R=70$ nm, two values of energy gap $\Delta=0, 50$ meV, and for three values of magnetic flux (a,d): $\eta=0$, (b,e): $\eta=0.1$, (c,f): $\eta=2.5$. The dotted green lines represent the electronic states $\tau=-1$, and the magenta lines are for the electronic states $\tau=1$.}
	\label{fig2art}
\end{figure}

We numerically analyze \eqref{e:15} for a fixed radius $R=70$ nm of the GMQD. Fig. \ref{fig2art} shows the energy spectrum versus the magnetic field $B$ for five values of the angular momentum $m = 0, \pm1,\cdots,\pm4$, three magnetic flux values: (a,d): $\eta=0$, (b,e): $\eta=0.1$,  (c,f): $\eta=2.5$, and two values of the energy gap $\Delta=0, 50$ meV. The dashed green curve and the magenta curve, respectively, reflect the energy levels that correspond to the $K'$ and $K$ valleys. For $\Delta=0$ and $\eta=0$, Fig. \ref{fig2art}a shows that the Landau level with zero energy is formed by electronic states with positive ($\tau=1$) and negative ($\tau=-1$) energies. For low values of $B$, we observe the existence of several electronic states with degenerate energies corresponding to all numbers of angular momentum $m$. Subsequently, with the increase in $B$, the creation of Landau levels becomes apparent. For $B=0$, the energy difference between the first positive electronic state and the first negative electronic state is approximately 8 meV, consistent with results from \cite{schnez2008analytic}. In Figs. \ref{fig2art}b and \ref{fig2art}c for $\eta=0.1$ and $2.5$, respectively, we observe an increase in the energy spectrum with magnetic flux, forming new Landau levels, in agreement with previous literature \cite{Belokda23, bouhlal23}. Additionally, there is an increase in the energy gap between the positive energy levels of holes and those of electrons. In Figs. \ref{fig2art}d, \ref{fig2art}e, and \ref{fig2art}f for $\Delta=50$ meV, the number of Landau levels formed is improved compared to the case of $\Delta=0$ meV. Furthermore, $\Delta=50$ meV leads to a clear widening of the bands, separating the Landau levels of positive and negative energies.

\begin{figure}[H]
	\centering
	\includegraphics[scale=0.46]{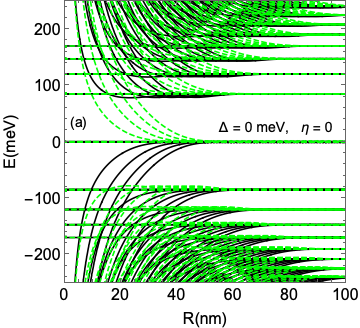}\includegraphics[scale=0.46]{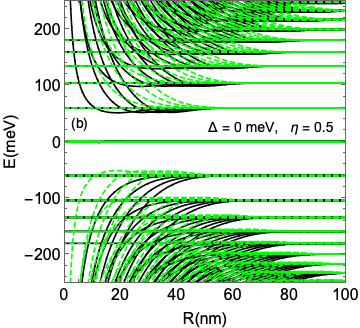}\includegraphics[scale=0.46]{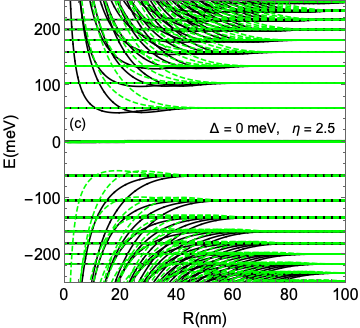}
	\includegraphics[scale=0.46]{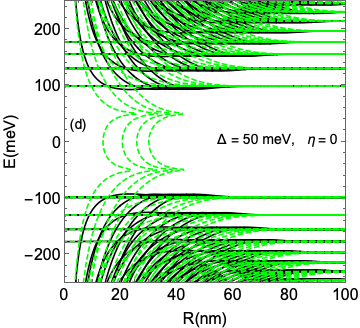}\includegraphics[scale=0.46]{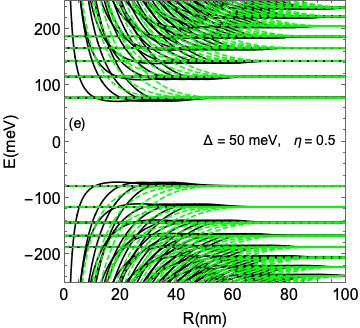}\includegraphics[scale=0.46]{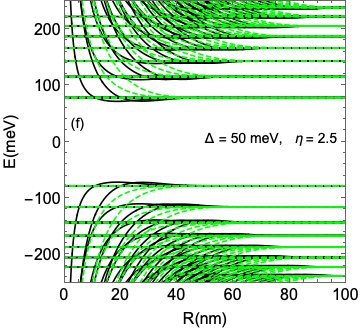}
	\caption{(color online) Energy spectrum $E$ as a function of the radius $R$ for $B=5.5$ T,  $m = 0, \pm1,\cdots,\pm4$, $\Delta=0, 50$ meV and three values of magnetic flux (a,d): $\eta=0$, (b,e): $\eta=0.5$, (c,f): $\eta=2.5$. The electronic states $\tau=-1$ are represented by the dotted green lines, and those $\tau=1$ by the black lines.}
	\label{fig3art}
\end{figure}

In Fig. \ref{fig3art}, we plot the energy spectrum versus the radius $R$ of the GMQD for the angular momentum $m = 0, \pm1,\cdots,\pm4$ and magnetic field $B = 5.5$ T. The magnetic flux $\eta$ and  energy gap $\Delta$ vary as follows: for each column, $\eta$ takes values $0, 0.1, 2.5$, and for each row, the  energy gap $\Delta$ is set to $0,50$ meV. The black and green dashed curves, respectively, represent the energy levels corresponding to the $K$ and $K'$ valleys. 
For $\Delta=0$ and $\eta=0$ in Fig. \ref{fig3art}a, we observe that the electronic states corresponding to all values of $m$ contribute to the energy spectrum. These states exhibit decreasing waviness for smaller values of $R$ and approach almost linear behavior as $R$ increases. Furthermore, it is evident that higher values of $R$ lead to a more pronounced decrease in these electronic states until they converge to form the Landau levels. Consequently, for low values of $R$, the positive and negative energy electronic states combine to generate the Landau level with zero energy. As $R$ increases, we observe that the energetic states closest to the zero-energy Landau level decrease until they converge at $E=0$. We choose $\eta=0.1$ in Figs. \ref{fig3art}b and 2.5 in \ref{fig3art}c to show that the formation of Landau levels with the increase of $R$ is still evident, with the exception that, in this instance, the number of Landau levels that develop rises with increasing $\eta$.
Moreover, we observe that as $\eta$ increases, the number of electronic states forming the Landau levels decreases. In comparison to Figs. \ref{fig3art}a, \ref{fig3art}b, and \ref{fig3art}c, it is evident that the introduction of an energy gap ($\Delta=50$ meV) in Figs. \ref{fig3art}d, \ref{fig3art}e, and \ref{fig3art}f results in an expansion of the forbidden band width, separating Landau levels with negative energy (electron) and the corresponding Landau levels with positive energy (hole). This implies the potential to regulate the movement of electrons from the valence band to the conduction band.

\begin{figure}[H]
	\includegraphics[scale=0.46]{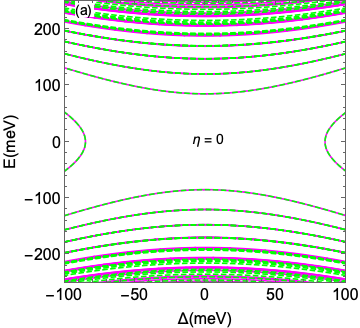}\includegraphics[scale=0.46]{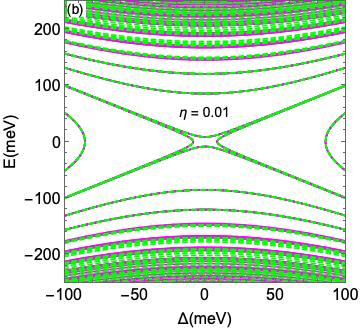}\includegraphics[scale=0.46]{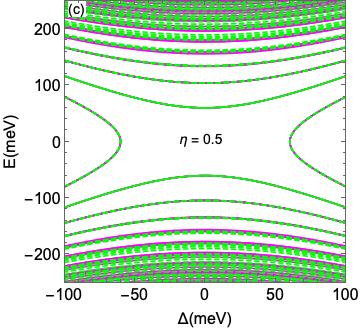}
	\caption{(color online) Energy spectrum $E$ as a function of the energy gap $\Delta$ for $R=70$ nm, $B=5.5$ T,  $m = 0, \pm1,\cdots,\pm4$, and   (a): $\eta=0$, (b): $\eta=0.01$, (c):  $\eta=0.5$. The dotted green lines indicate the electronic states $\tau=-1$, whereas the magenta lines represent the electronic states $\tau=1$.}
	\label{fig4art}
\end{figure}

Additionally, we can analyze the energy spectrum as a function of the energy gap $\Delta$ for all angular momentum values ranging from -4 to 4. Here we choose the radius of $R=70$ nm, magnetic field $B=5.5$ T, and three values of magnetic flux: $\eta=0, 0.01, 0.5$.  Fig. \ref{fig4art}, illustrates the results where the energy levels associated with the $K'$ and $K$ valleys are represented by the dashed green and magenta curves, respectively. In Fig. \ref{fig4art}a, we observe a parabolic behavior of energy levels, with a minimum corresponding to $\Delta=0$ meV. For energies $E > 100$ meV, the majority of the energy levels exhibit a horizontal parabolic form, while below this energy value, some energy levels adopt a vertical parabolic form. For non-null values of $\eta$, we observe the generation of new parabolic energy levels. The energy band separating these new levels becomes considerably wider as the magnetic flux increases, as depicted in Figs. \ref{fig4art}b and \ref{fig4art}c.

\begin{figure}[H]
	\centering
	\includegraphics[scale=0.45]{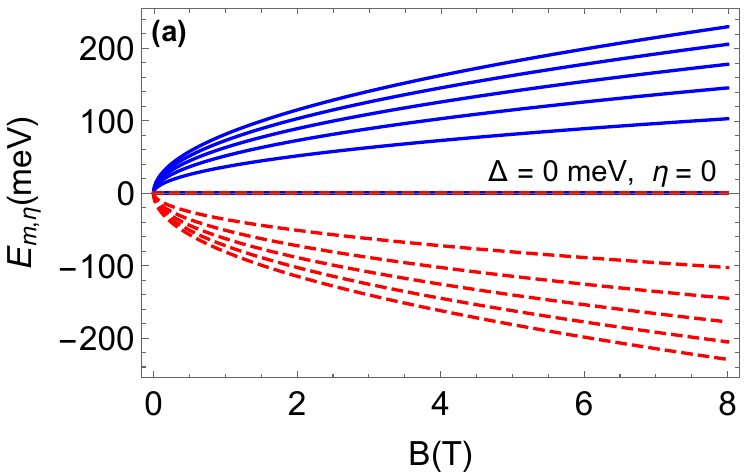}  \includegraphics[scale=0.45]{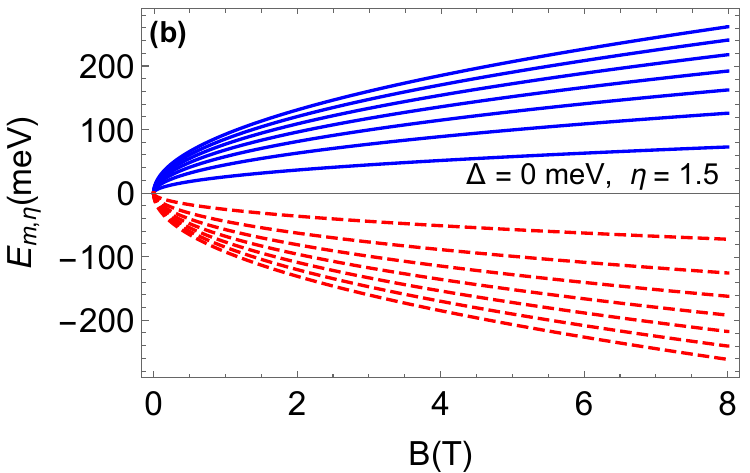}  \includegraphics[scale=0.45]{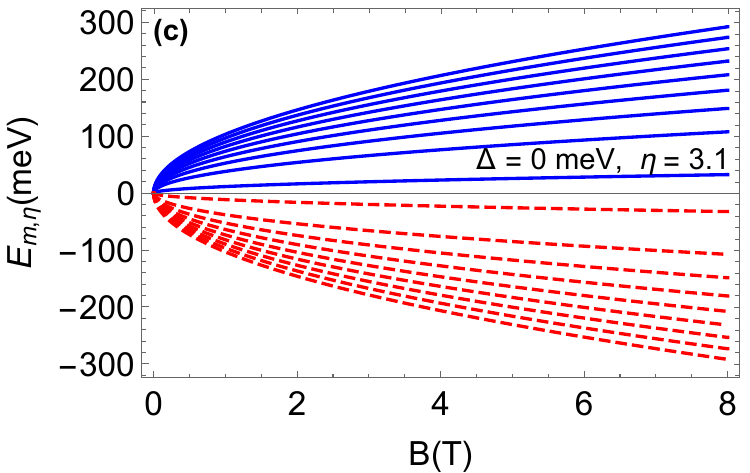}
	\includegraphics[scale=0.45]{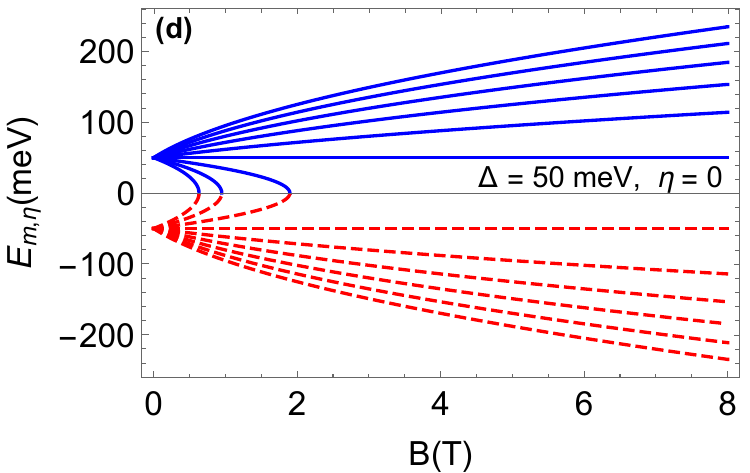}  \includegraphics[scale=0.45]{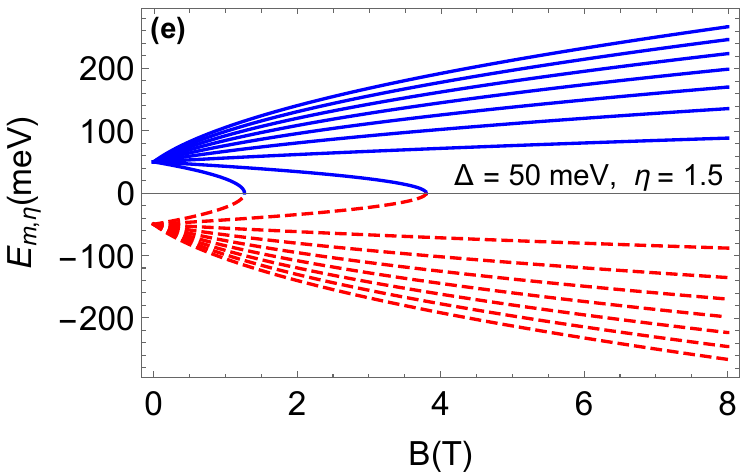}  \includegraphics[scale=0.45]{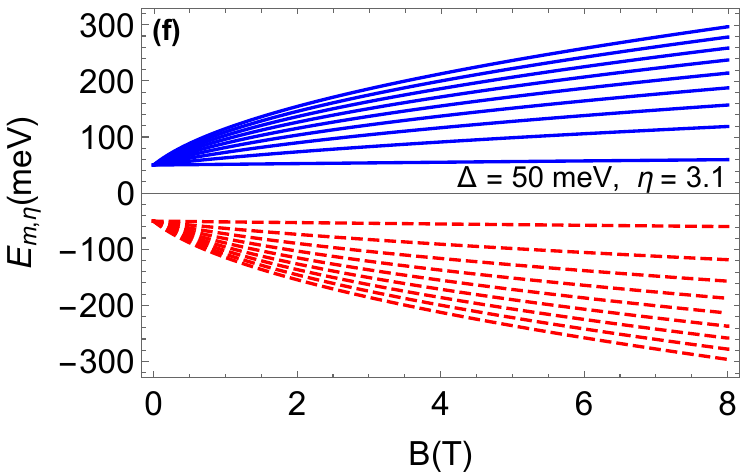}
	\caption{(color online) Energy $E_{m,\eta}$ as a function of magnetic field  $B$ for $m = 0, \pm1,\cdots ,\pm4$, $\Delta=0, 50$ meV, and (a,d): $\eta=0$, (b,e): $\eta=1.5$, (c,f): $\eta=3.1$. The blue lines represent the energy levels $K$ , and the red dashed lines represent the energy levels $K'$.}
	\label{fig5art}
\end{figure}

Next, we analyze the numerical behavior of \eqref{e:18}, which represents the eigenenergies of the system under infinite-mass boundary conditions. For any value of angular momentum $m = -4,\cdots,4$, the depiction of the energy spectrum $E_{m,\eta}$ as a function of the magnetic field  $B$ is displayed in Fig.\ref{fig5art}. The magnetic flux $\eta$ takes values of $0, 1.5, 3.1$ for each column, and the  energy gap is $\Delta=0$ meV for the first line and $\Delta=50$ meV for the second line. The blue lines represent the energy levels corresponding to the $K$ valley, while the red dotted lines represent those corresponding to the $K'$ valley. Fig. \ref{fig5art}a clearly shows only detection of the energy levels corresponding to the first five positive states of the angular momentum $m$ in the absence of an energy gap and magnetic flux. These are the well-known Landau levels of magnetically-induced graphene.
The level with energy $E=0$ meV is formed by combining the various energy levels that correspond to negative values of the angular momentum $m$. When we increase $\eta$, we see that the energy levels corresponding to the negative states of the angular momentum $m$ start to be visible, until all four states corresponding to $m=-1, -2, -3, -4$ appear at a value $\eta=3.1$, as shown in Figs. \ref{fig5art}b and \ref{fig5art}c.
For $\Delta=50$ meV in Figs. \ref{fig5art}d and \ref{fig5art}e, we observe other energy states with negative values of the angular momentum $m$ corresponding to additional energy levels with a vertical parabolic form.  When the magnetic flux increases, we note that the energy behavior changes with the formation of a band gap of energy $\Delta=100$ meV separating the conduction band and the valence band, as clearly seen in Fig. \ref{fig5art}f.


\begin{figure}[H]
	\centering
	\includegraphics[scale=0.45]{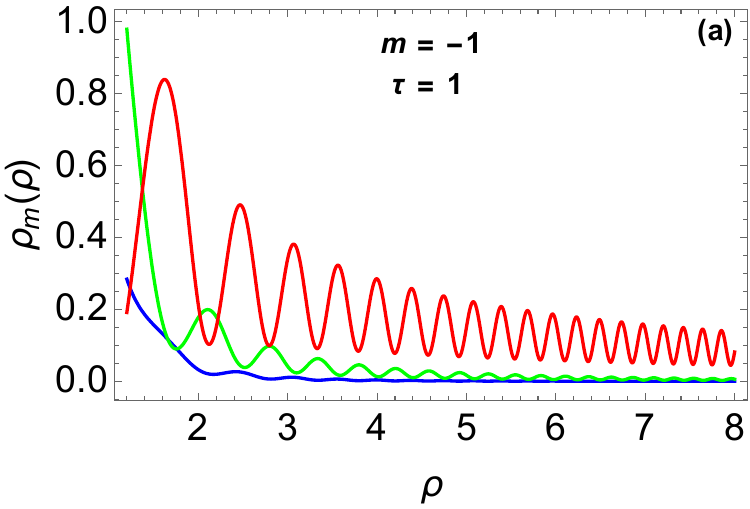}  \includegraphics[scale=0.45]{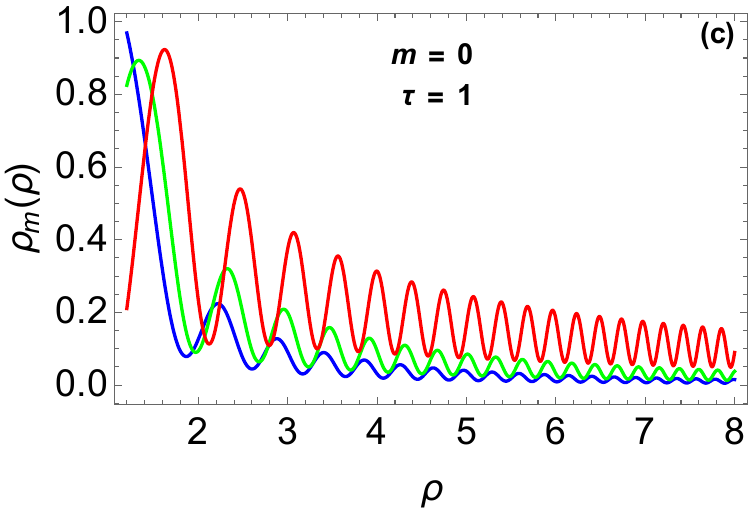}  \includegraphics[scale=0.45]{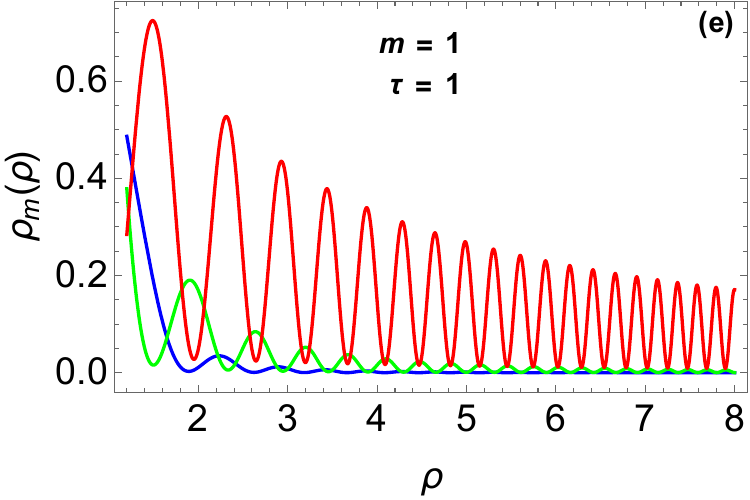}
	\includegraphics[scale=0.45]{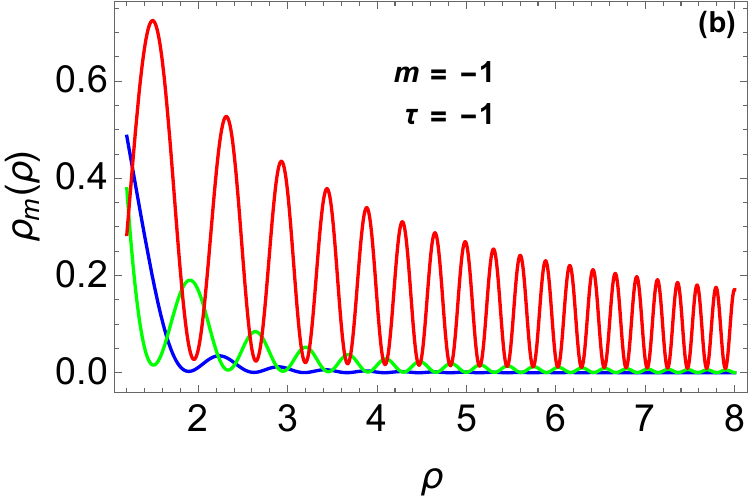}  \includegraphics[scale=0.45]{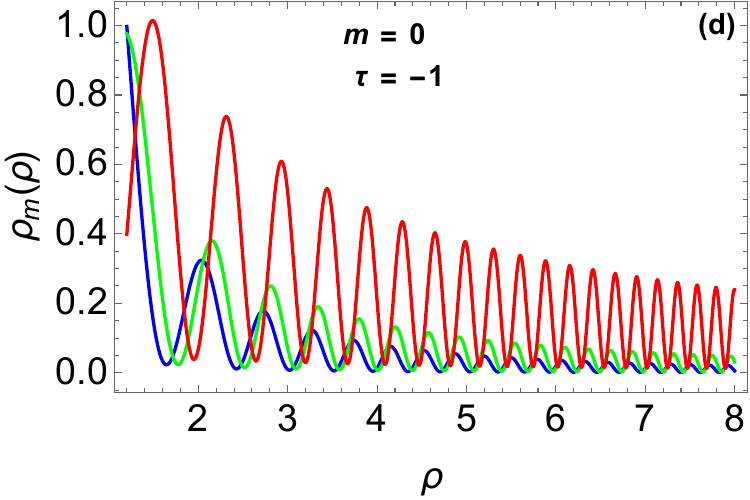}  \includegraphics[scale=0.45]{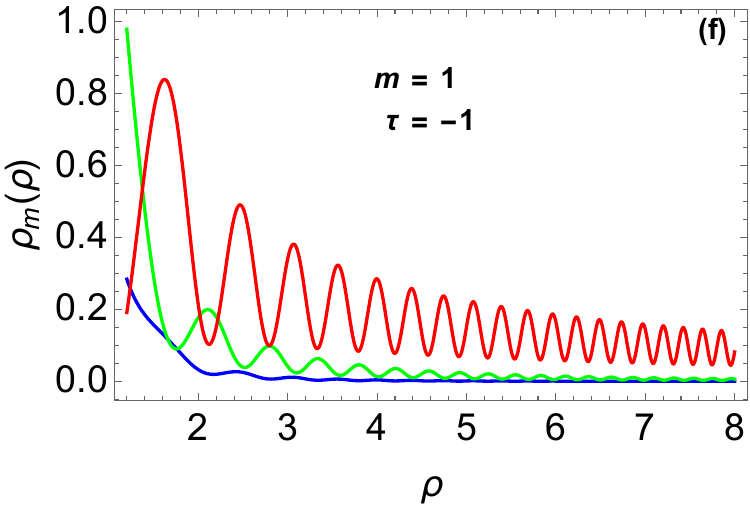}
	\caption{(color online) Radial probability $\rho_m(\rho)$ for the two valleys $K$ and $K'$ as a function of the normalized radius $\rho$ for $B=15$ T, $E=100$ meV, $\Delta=0$, and  three values of the angular momentum number (a, b): $m=-1$ , (c, d): $m=0$, (e, f): $m=1$. The magnetic flux values considered are  $\eta=0$ (blue line), 0.5 (green line), and 1 (red line).}
	\label{fig6art}
\end{figure}

Fig. \ref{fig6art} shows the radial probability $\rho_m(\rho)$ plotted against the normalized radius $\rho=\frac{R}{l_B}$ for the two valleys, $K(\tau=1)$ and $K'(\tau=-1)$, with an incident energy $E = 100$ meV, a magnetic field  $B = 15$ T, an energy gap $\Delta=0$, and the angular momentum $m = -1, 0, +1$. We vary the magnetic flux values as follows: $\eta=0$ (blue line), $\eta=0.5$ (green line), and $\eta=1$ (red line).
The radial probability $\rho_m(\rho)$ peaks at the center of the quantum dot ($\rho$ = 0) when there is no magnetic flux (blue line).
With increasing $\rho$, the radial probability gradually decreases until it stabilizes at a minimum value close to 0 when the angular momentum takes the values $m = -1$ and $m = 1$, as depicted in Figs. \ref{fig6art}a, \ref{fig6art}b, \ref{fig6art}e, and \ref{fig6art}f. 
It is worth noting that our results are consistent with those presented in \cite{bouhlal23}.
Nevertheless, when angular momentum is zero $(m=0)$, the radial probability experiences a swift decline as $\rho$ increases. This is evident in Figs. \ref{fig6art}c and \ref{fig6art}d, where resonance peaks are noticeable and become damped as $\rho$ is elevated.
This becomes apparent following the initial peak, which approaches a value close to 1. Generally, the radial probability attains a maximum close to the center of the quantum dot when the magnetic flux is adjusted towards non-zero values (indicated by the green and red lines). However, in this specific scenario, we note that the radial probability exhibits oscillatory patterns with multiple maxima. These maxima correspond to the confined quasi-bound states or trapping states of electrons within the GMQD as $\rho$ increases.



\begin{figure}[H]
	\centering
	\includegraphics[scale=0.45]{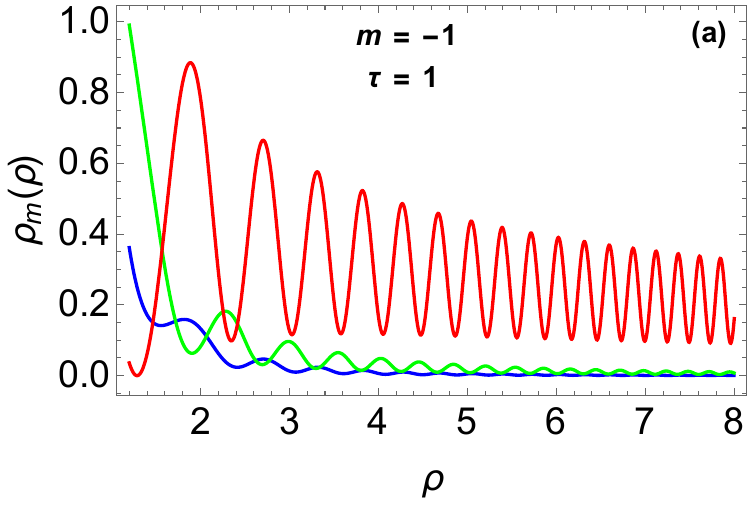}  \includegraphics[scale=0.45]{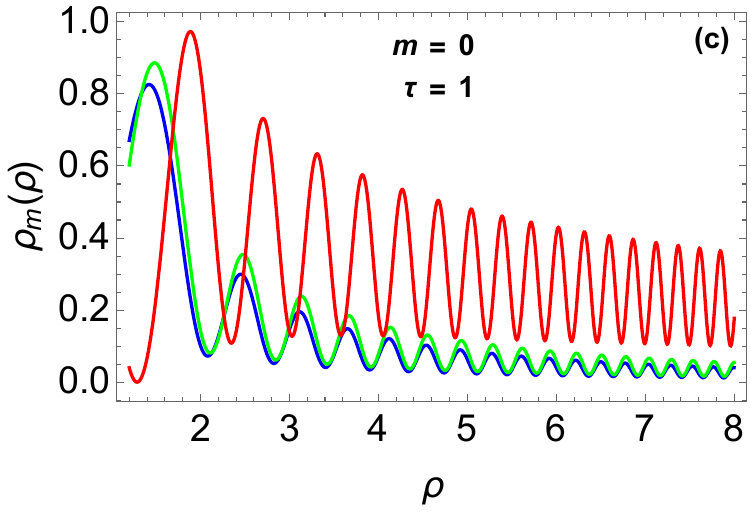}  \includegraphics[scale=0.45]{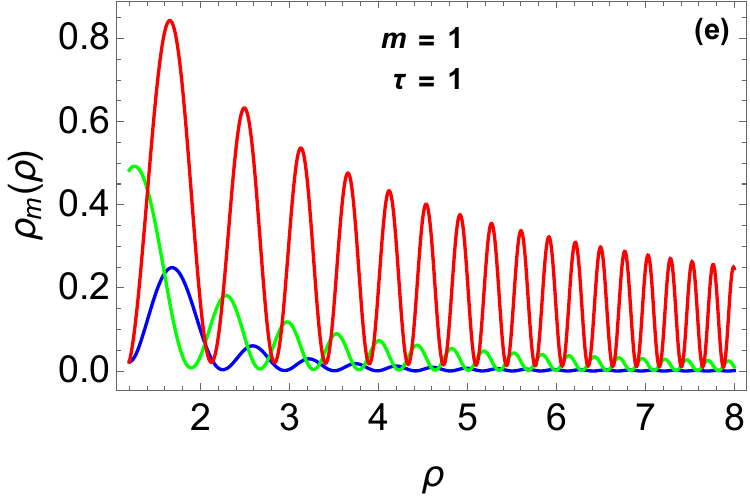}
	\includegraphics[scale=0.45]{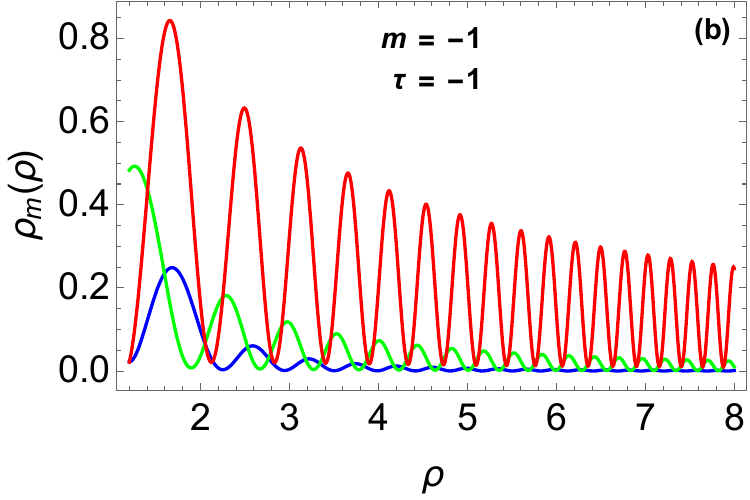}  \includegraphics[scale=0.45]{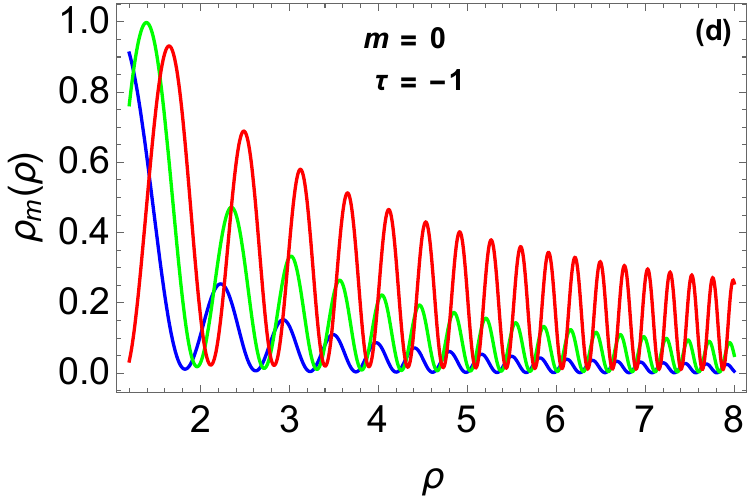}  \includegraphics[scale=0.45]{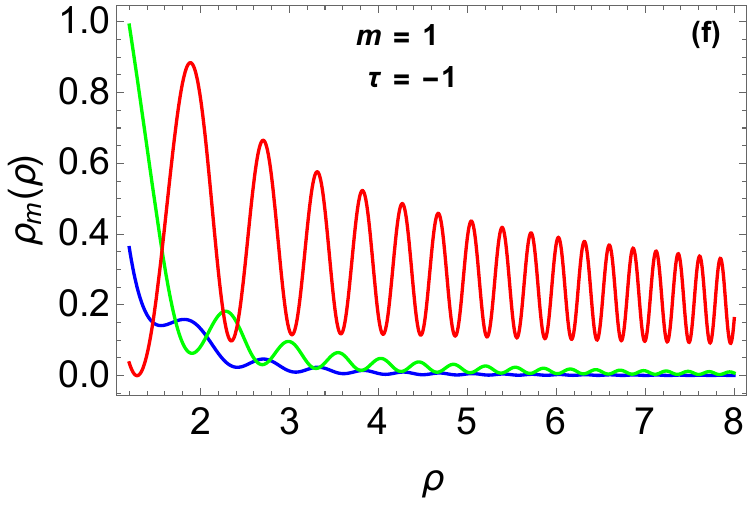}
	\caption{(color online) Radial probability $\rho_m(\rho)$ for the two valleys $K$ and $K'$ as a function of the normalized radius $\rho$ for $B=15$ T, $E=100$ meV, $\Delta=30$ meV, and  three values of the angular momentum number (a, b): $m=-1$ , (c, d): $m=0$, (e, f): $m=1$. The magnetic flux values considered are  $\eta=0$ (blue line), 0.5 (green line), and 1 (red line).}
	\label{fig7art}
\end{figure}

In Fig. \ref{fig7art}, incorporating an energy gap of $\Delta=30$ meV, we consistently observe maximum radial probability near the center of the quantum dot. Moving away from the center, the radial probability exhibits an oscillatory decrease, reaching multiple maxima during this decline. These maxima signify quasi-bound states or electron-trapping states within the quantum dot. While these results share similarities with those in Fig. \ref{fig6art}, the introduction of the energy gap in this case has a positive impact on the amplitudes of the resonance peaks. Additionally, we note that the number of maximum values attained by the radial probability is fewer than what was observed in the absence of energy gap  ($\Delta=0$).
 
\section{Conclusion} \label{cc}
The present work introduces a theoretical framework for studying the behavior of Dirac fermions in graphene quantum dots (GMQD) under the influence of a magnetic field, a magnetic flux, and an inhomogeneous energy gap. The study focuses on the effects of these parameters on the energy spectrum exhibited by the present system. We analytically solved the Dirac equation and then determined the eigenspinors using a mathematical approach. By employing the infinite mass boundary condition, we derived an expression governing the corresponding energy spectrum. This involves a dependence on physical parameters such as magnetic flux $\phi$, angular momentum $m$,  radius $R$ of GMQD, energy gap $\Delta$, and magnetic field $B$.


In our numerical investigation of the energy spectrum of the GMQD, we explored various values of the system's physical properties based on theoretical predictions. Our observations revealed a notable enhancement in the energy spectrum within the GMQD as the magnetic flux increased, accompanied by a slight expansion between energy levels. Furthermore, we noted that the gap between energy levels occupied by electrons and the corresponding levels occupied by holes widened with an increase in the energy gap. As the magnetic flux rose, we observed the emergence of new levels associated with negative values of angular momentum in the region where electron energies are governed by confinement in the Landau levels ($R/\l_B$$\to$$\infty$). Additionally, we identified the formation of a band gap between the valence and conduction bands with an increase in the energy gap.

The comprehension of radial probability is essential for grasping the distribution of localized electrons and holes as well as determining the positions of bound states. In our investigation, we observed that the presence of magnetic flux within the GMQD plays a pivotal role in regulating electron confinement and extending the duration of quasi-bound states in the GMQD. Therefore, it can be inferred that introducing magnetic flux and an energy gap is instrumental in governing the mobility of electrons within GMQDs, consequently influencing the electrical characteristics of the GMQDs.

\end{document}